\documentclass{article}
\usepackage{jheppub} 
\usepackage{graphicx}
\usepackage{amsmath,amsthm, amssymb, amsfonts}
\usepackage{braket}
\usepackage{natbib}
\usepackage{xcolor}

\newcommand{\Y}{\mathcal{Y}}
\newcommand{\OO}{\mathcal{O}}
\newcommand{\I}{{lk\alpha}}
\newcommand{\D}{{\Delta_{l}}}
\newcommand{\LA}{{L}}
\newcommand{\dd}{\partial}
\newcommand{\R}{\mathbb{R}}
\newcommand{\Z}{\mathbb{Z}}
\newcommand{\M}{\mathcal{M}}
\newcommand{\N}{\mathcal{N}}
\newcommand{\ha}{{\hat {\delta}_{ij}}}
\newcommand{\ch}{{\check {\delta}_{ij}}}
\newcommand{\sha}{{\hat {s}_{ij}}}
\newcommand{\sh}{{\check{s}_{ij}}}
\newcommand{\slap}{{\nabla^2_{S^p}}}

\title{Kinematically Unrestricted Flat Space Scattering from AdS\textsubscript{d+1} $\times$ S\textsuperscript{p}/CFT\textsubscript{d}}
\author{Bhanu Narra} 

\affiliation{Pembroke College, University of Oxford, \\ St. Aldates, Oxford, OX1 1DW, UK }
\emailAdd{bhanu.narra@pmb.ox.ac.uk}

\abstract{We construct scattering states for scalar fields with generic kinematics from CFT operators in the flat space limit of AdS/CFT. The states can have momentum along the compact $S^p$ directions of $AdS_{d+1}\times S^p$, which is achieved by scaling to large R-symmetry charge and choosing a particular parameterization for the R-symmetry polarization. These states allow for the construction of an unrestricted S-matrix from the flat space limit of CFT correlation functions, which we check for various tree level interactions. We also describe a generalization of the Mellin space formalism which incorporates the compact momentum, and we discuss some possible implications for flat space holography.}

\begin{document}

\maketitle

\section{Introduction}

Holography in AdS has been well established \cite{Maldacena_1999,witten1998antisitterspaceholography,Gubser_1998}. A natural way to attempt to understand holography in spacetimes different from AdS is by taking the curvature scale of AdS to infinity, $\LA\to \infty$, wherein $AdS_{d+1}$ flattens to $d+1$ dimensional Minkowski space $\R^{d,1}$. One can hope that this limit could yield a holographic understanding of flat spacetime. Originating with \cite{polchinski1999smatricesadsspacetime, Susskind_1999, Giddings:1999qu, Giddings_2000}, several works have attempted to extract the flat space S-matrix from CFT correlation functions in this limit.

In many of the known cases of AdS/CFT, for instance those coming from Freund-Rubin solutions of supergravity \cite{FREUND1980233}, one also has compact extra dimensions (in the simplest cases $S^p$) which also decompactify in the flat space limit. In these Freund-Rubin solutions, the radius of the sphere is proportional to the AdS curvature scale $R = \chi \LA$. For example, $\chi = 1$ for $AdS_5 \times S^5$ and $\chi = 2$ for $AdS_4 \times S^7$. As such, in the flat space limit we do not just obtain $\R^{d,1}$, but rather $\R^{d+p,1}$. If the AdS distance conjecture \cite{L_st_2019} is true, then one might expect that this is not merely a consequence of supersymmetry and that the flat space limit is generally accompanied by the decompactification of some compact space. The presence of $S^p$ gives rise to a tower of KK modes which appear as representations of $SO(p+1)$ in the boundary theory; we will call this $SO(p+1)$ symmetry the R-symmetry, even though it may not necessarily be part of a supersymmetry algebra.

Many works on the flat space limit of AdS/CFT focus primarily on the noncompact AdS directions. As such, the kinematics of the external states in the S-matrix are restricted to a subplane $\R^{d,1}\subset \R^{d+p,1}$ in the flat space limit. For $n$-point functions with $n-1 \le d+1$ this is generally not an issue, since the external states can be linearly independent, and one does not miss out on any information about the full $D=d+p+1$ dimensional S-matrix (at least for scalar fields). However, for higher point functions, it is possible that the extra linear dependence from the condition that all external momenta lie in $\R^{d,1}$ may lead to missing terms in the generic S-matrix; even if this does not occur, it seems desirable to remove the restriction so that we can obtain a more generic flat space observable. The kinematic restriction also has implications for attempts to obtain flat space holography via a limit of AdS/CFT. For instance, we might expect the dual of $\R^{d+p,1}$ to be a $d+p$ dimensional Carrollian CFT. However, some attempts at taking flat space limits of $AdS_{d+1}\times S^p$/CFT$_d$ via taking Carrollian limits of the boundary give rise to a $d$ dimensional CFT instead of a $d+p$ dimensional CFT \cite{Lipstein:2025jfj, bagchi2026doublyscaledplanarcaln}, see also \cite{Fontanella_2026}. Thus $p$ dimensions of the boundary CFT naïvely seem to be `missing'.

In this article, we describe how to obtain generic kinematics for the $D$ dimensional S-matrix. Our main result is a formula for the unrestricted $D$ dimensional $n$ particle S-matrix for scalar fields in terms of a CFT $n$ point function (\ref{generic S matrix}). The idea is that by choosing the R-symmetry polarizations $y_I$ in an appropriate way ($y_{qI}=(i\hat q, 1)$) and scaling the R-symmetry charge $l$ of the representation (the angular momentum on $S^p$) to infinity, keeping the ratio $q=l/R$ fixed, we can create a bulk scattering state which has nonzero momentum $\vec q$ in the compact directions in the flat space limit. Our approach is primarily derived from the global smearing formalism established in \cite{fitzpatrick2011scatteringstatesadscft, Fitzpatrick_2012, Li_2021}. The large charge limit has also been studied and discussed primarily in the context of CFT four point functions, e.g. in \cite{Aprile_2020,Aprile_2021,Aprile_2023, Chester_2018}, though a general formula like (\ref{generic S matrix}) does not seem to appear in the literature. This formula and the related formula for the generic scattering state (\ref{eq: momentum state}) may also be of interest to Carrollian holography; one possible interpretation is that in the flat space limit the R-symmetry algebra gains an interpretation as a spacetime algebra in the boundary Carrollian CFT, thereby restoring the $p$ missing dimensions above.

A recent paper \cite{Eden:2026ulx} has also addressed the decompactification issue for cases with $p=d+1$ and $\chi = 1$, i.e. the case when the AdS and sphere radii are equivalent. Their approach uses a resummed master propagator which efficiently captures the dynamics of the various KK modes. On the boundary, they define an operator which also resums all the KK modes. Their prescription similarly allows them to extract S-matrices with generic kinematics from CFT correlation functions.

While there is overlap, our approach differs in a few ways from \cite{Eden:2026ulx}. Firstly, their master propagator relies on the conformal flatness of $AdS_{d+1}\times S^{d+1}$ when $\chi =1$ (which also gives rise to special symmetries in $\mathcal{N}=4$ SYM correlation functions \cite{Caron_Huot_2019}), and so it does not seem to exist in cases where $\chi \neq 1$. As such their approach may not directly apply to $AdS_{4}\times S^7$, for example. Our approach can accommodate $\chi\neq 1$, but it is restricted to ($D$ dimensional) scalar fields for now; M-theory in $\R^{10,1}$ has no scalar fields, so our approach does not quite apply to $AdS_{4}\times S^7$ either, though the needed generalization to higher spin fields seems like a clearer future target. In addition, in some sense our approach is `more intrinsic to the boundary': we give a direct formula to go from a CFT operator to a generic bulk scattering state (\ref{eq: momentum state}). On the other hand, their approach seems like it could be more useful for computations than ours, since our formula requires a large charge limit of a general CFT correlation function, while their master propagator has a very simple expression. Our formula also requires a smearing which complicates it further. Some of the finer details also appear to differ, such as the parameterization of the null polarization. 

We will also construct a double Mellin representation to obtain flat space S-matrices. This is in line with the standard Mellin space formalism of \cite{Penedones_2011,Fitzpatrick_2011}, which relates the S-matrix to the Mellin transform of a CFT correlation function. The Mellin space variables $\delta_{ij}$ are ultimately related to the Mandelstam invariants $s_{ij}=-(p_i+p_j)^2$. It has been anticipated before that incorporating the sphere polarizations $y_I$ via a Mellin sum would be able to account for the contributions to the Mandelstam invariants coming from momentum in the compact directions \cite{Rastelli:2016nze,Rastelli_2018,Caron_Huot_2019,Aprile_2021,Aprile_2023}, again in the context of supersymmetric four point functions. We are able to show this explicitly in the more general case by doing a saddle point analysis akin to \cite{Fitzpatrick_2012, Li_2021}. 

In section \ref{Sec: obtaining S matrix}, we derive our formula for scattering states with generic momentum and the corresponding S-matrix. In section \ref{Sec: Checks}, we perform some basic consistency checks of the proposal for various scalar field interactions at tree level. In section \ref{Sec: Double Mellin representation}, we describe the double Mellin representation. We conclude in section \ref{Sec: Conclusion}.

\section{Extracting the $D$-dimensional S-matrix}\label{Sec: obtaining S matrix}

Let the total spacetime dimension be $D$, the noncompact dimension $d+1$, and the compact dimension $p$, so $D = d+p+1$. We denote the harmonics on $S^p$ as $Y^p_\I$,\footnote{The index $\I$ is explained in appendix \ref{Appendix A}.} and we assume the sphere radius is proportional to the AdS curvature length $R=\chi \LA$. Consider a $D$ dimensional massless scalar field $\Phi$, which can be expanded as\footnote{We use uppercase latin indices for full spacetime indices, lowercase latin for compact space indices, and greek for noncompact indices.} 
\begin{equation}
    \Phi (x^M) = \sum_{\I} \phi^{\I} (x^\mu )Y^p_{\I}(x^m).
\end{equation}
Each KK mode $\phi^{\I}(x^\mu)$ has mass $m_l^2=l(l+p-1)/(\chi \LA)^2$. Each mode also has a dual operator $\OO^{\I}(x)$ of weight $\D $ satisfying $\D(\D -d) =m_l^2\LA^2$, i.e. 
\begin{equation}
    \D = \frac{d+ \sqrt{d^2+4l(l+p-1)/\chi^2}}{2}. 
\end{equation}
The KK modes $\phi ^\I$ are in traceless symmetric rank $l$ tensor representations of the R-symmetry.

We will use global coordinates on $AdS_{d+1} \times S^p$, 
\begin{equation}
    ds^2=\frac{\LA^2}{\cos^2(\rho)}(-d\tau ^2 + d\rho^2 + \sin^2(\rho)d\Omega_{d-1}^2) +(\chi \LA)^2(d\theta^2 + \sin^2(\theta) d\Omega_{p-1}^2). 
\end{equation}
In these coordinates, the $D$ dimensional field can be quantized as 
\begin{equation}
    \Phi (x^M) = \sum _{nJ m  \I} \left( \varphi^\I_{nJm} (x^M)a_{nJm\I} + \varphi^{*\I}_{nJm} (x^M)a^\dagger_{nJm\I}\right),
\end{equation}
where \cite{fitzpatrick2011scatteringstatesadscft, Li_2021}
\begin{equation}
    \varphi ^\I_{nJm}=e^{iE_{nJl} \tau } R_{nJl}(\rho)\Y_{Jm}(\hat \rho )Y_\I( x^m).
\end{equation}
Here,
\begin{align}
    E_{nJl} &= \D +2n +J,
    \\R_{nJl}(\rho)  &= \frac{1}{N_{nl J}} \sin^J(\rho)\cos^{\D}(\rho) {}_2F_1\left(-n, \D+J+n, J+\frac d 2 , \sin^2(\rho)\right), \\ N_{nl J}&=\sqrt{\frac{n!\Gamma(J+ \frac d 2)^2\Gamma(\D+n-\frac{d-2}{2})\LA^{d-1}}{\Gamma(n+J+\frac d 2)\Gamma(\D+n+J)}},
\end{align}
and $\Y_{Jm}$ are spherical harmonics on $S^{d-1}$.

Since AdS/CFT implies that the Fock space of the bulk gravitational theory is isomorphic to the Fock space of the CFT in this regime \cite{witten1998antisitterspaceholography}, the boundary operator dual to $\phi^\I$ can be quantized with the same modes \cite{fitzpatrick2011scatteringstatesadscft}:
\begin{equation}
    \OO ^\I(\hat \rho , \tau) = \sum {N_{nJl}^\OO}\left(e^{iE_{nJl}\tau} \Y_{Jm}(\hat \rho) a_{nJm\I} + e^{-iE_{nJl}\tau} \Y^*_{Jm}(\hat \rho) a^\dagger_{nJm\I}\right),
\end{equation}
with 
\begin{equation}
    N_{nJl}^\OO = \sqrt{\frac{\Gamma(1+\D-\frac d 2 +n)\Gamma(\D+J+n)}{\Gamma(1+\D-\frac d 2 )^2\Gamma(1+n)\Gamma(\frac d 2 + J +n)}}.
\end{equation}
This allows a boundary construction of the bulk creation operator including R-symmetry indices, 
\begin{equation}
    a^\dagger _{nJm\I} = \int \frac{d\tau}\pi d^{d-1}\hat x e^{iE_{nJl}\tau}\frac{\Y_{Jm}(\hat x)}{N_{nJl}^\OO}\OO^\I (\hat x, \tau).
\end{equation}
To obtain the desired S-matrix, we would like a sum of these modes which, in the flat space limit, reduces to a plane wave. A useful result is obtained in appendix \ref{Appendix A}:
\begin{equation}\label{flat limit of spherical harmonics}
    Y^p_{lk\alpha}(\theta, \Omega_{p-1}) \underset{\substack{R\to \infty \\ l\to \infty \\ l/R=q }}{\longrightarrow} \sqrt{\frac{q^{p-1}}{R}}\frac{J_{k+\nu} (qr)}{(qr)^{\nu}} Y^{p-1}_{k\alpha}(\Omega_{p-1}), 
\end{equation}
where  $r = R\theta$. This gives a representation of the spherical harmonic as a function in spherical coordinates on $\R^p$ in the flat space limit. To obtain a plane wave, we simply use the plane wave expansion 
\begin{equation}\label{plane wave explansion}
    e^{i\vec q \cdot \vec r} = (2\pi)^{p/2}\sum _{k=0}^\infty \sum _{\alpha =1}^{N(k, p-1)}i^k \frac{J_{k+\nu}(qr)}{(qr)^\nu} Y^{*p-1}_{k\alpha}(\hat q ) Y^{p-1}_{k\alpha}(\hat r),
\end{equation}
with $N(k, p-1) = \frac{(2k+p-2)(k+p-3)!}{k!(p-2)!}$ denoting the degeneracy in the index $\alpha$. The factor $Y^{*p-1}_{k\alpha}(\hat q )$ will allow us to determine the direction that the particle travels in the compact dimension, and the angular momentum $l/R$ becomes the magnitude of the rectilinear momentum $q$. To obtain a fully generic plane wave we combine with the noncompact directions \cite{fitzpatrick2011scatteringstatesadscft}. We have 
\begin{equation}
    e^{ip_\mu x^\mu + i q_m x^m} = \sum _{Jmk\alpha}(2\pi)^{p/2}i^{k+J}\frac{J_{k+\nu}(qr)}{(qr)^\nu} Y^{*p-1}_{k\alpha}(\hat q ) Y^{p-1}_{k\alpha}(\hat x^m)j_J(|\vec p|\rho)\Y_{Jm}^*(\hat p)\Y_{Jm}(\hat x^\mu),
\end{equation}
where $j_J(|\vec p|\rho)$ is a spherical Bessel function coming from the AdS radial direction. As we can see, taking a combination of modes of the form $Y^{*p-1}_{k\alpha}(\hat q)\Y_{Jm}(\hat p)a_{nJm\I}^\dagger$ and choosing $n$ and $l$ to obtain a well-defined energy and compact momentum will give us a bulk wavefunction which reduces to the plane wave $e^{ip_\mu x^\mu + i q_m x^m}$ in the flat space limit. More precisely, letting $\vec P = (\vec p, \vec q)$ be the total $D$ dimensional momentum, we should have
\begin{align}
    \ket{\vec P} &\stackrel{?}{=}\underset{\substack{\LA\to \infty \\ l\to \infty \\ l/\chi\LA=q }}{\lim} \sum _{Jm}\sum _{k=0}^l \sum _\alpha (2\pi)^{p/2}\sqrt{\frac{\chi \LA}{q^{p-1}}}Y^{*p-1}_{k\alpha}(\hat q)\Y^*_{Jm}(\hat p)a^{\dagger}_{nJmlk\alpha}\ket{0} \nonumber\\ &= \int \frac{d\tau}\pi d^{d-1}\hat x \underset{\substack{\LA\to \infty \\ l\to \infty \\ l/\chi\LA=q }}{\lim} \sum _{Jmk\alpha} (2\pi)^{p/2}\sqrt{\frac{\chi \LA}{q^{p-1}}}e^{iE\tau }\frac{1}{N_{nJl}^\OO}Y^{*p-1}_{k\alpha}(\hat q)\Y^*_{Jm}(\hat p)\mathcal{Y}_{Jm}(\hat x)\OO^{lka}(\hat x, \tau)\ket{0}.\nonumber
\end{align}
This can be simplified and using the limiting expression for the normalization obtained in \cite{fitzpatrick2011scatteringstatesadscft, Li_2021},
\begin{align}\label{momentum state 1}
     \ket{\vec P} &\stackrel{?}{=}\int dtd^{d-1}\hat x \underset{\substack{\LA\to \infty \\ l\to \infty \\ l/\chi\LA=q }}{\lim}\sum _{k\alpha} e^{i\omega t}\LA^{\frac{d-1}2-\D} (2\pi)^{p/2}\sqrt{\frac{\chi \LA}{q^{p-1}}} \xi_{\omega, \D} \nonumber \\ & \;\;\;\;\;\;\;\;\;\;\;\;\;\;\;\;\;\;\;\;\;\;\;\;\;\;\;\;\;\;\;\;\;\;\;\;\;\;\;\;\;\;\;\;\;\;\;\;\;\;\;\;\times |\vec p|^{1-\D}2^\D \pi^{\frac{d-2}2} \Gamma\left(1+\D -\frac d 2\right) \delta(\hat x- \hat p) Y^{*p-1}_{k\alpha}(\hat q)\OO^{lka}(\hat x, t)\ket{0} \nonumber \\ &= \int dt \underset{\substack{\LA\to \infty \\ l\to \infty \\ l/\chi\LA=q }}{\lim}\sum _{k\alpha} e^{i\omega t}\LA^{\frac{d-1}2-\D} (2\pi)^{p/2}\sqrt{\frac{\chi \LA}{q^{p-1}}} \xi_{\omega, \D}|\vec p|^{1-\D}2^\D \pi^{\frac{d-2}2} \Gamma\left(1+\D -\frac d 2\right)  Y^{*p-1}_{k\alpha}(\hat q)\OO^{lka}(\hat p, t)\ket{0},
\end{align}
with 
\begin{align}
    t&=\tau/L \nonumber \\n&=(E-\D-J)/2 \nonumber \\ E&=\sqrt{\vec p^2+\vec q^2}\LA = \omega \LA \nonumber \\
    \xi_{\omega, \D} &= \left( \frac{\omega \LA - \D}{\omega \LA + \D}\right)^{\omega \LA/2}e^\D \nonumber .
\end{align}
We have absorbed the phases of the plane wave expansion into the creation operators, as in \cite{fitzpatrick2011scatteringstatesadscft, Li_2021}. 

Unfortunately, this state is not quite correct as written. This is because for a fixed level $l$, the spherical harmonics satisfy a parity condition $Y_\I (-n) = (-1)^lY_\I (n).$ Thus, the above state produces a plane wave at the north \textit{and} south poles of $S^p$ in the flat space limit, which leads to factors of two being wrong in the S-matrix. To resolve this, we need to smear across a range of levels $\Delta l\ll R$ to produce a `bump function' at the north pole of the sphere (while retaining a well-defined momentum $q$), thereby eliminating the mirror image at the south pole. 

Before describing this smearing, it is convenient to change basis in the R-symmetry indices, which will provide a significantly simpler representation of the necessary plane wave. The R-symmetry content of CFT operators is usually written using an auxiliary $p+1$ dimensional null vector $y_I$,
\begin{equation}
    \OO^l(x, y) = \OO^{I_1 \dots I_l}y_{I_1}\dots y_{I_l}.
\end{equation}
Here, instead of using the indices $\I$, we use the fundamental index $I_i \in \{1, \dots, p+1\}$ of $SO(p+1)$. Spherical harmonics at level $l$ are given by traceless symmetric tensors of rank $l$ in these indices. Given such a tensor $C_{I_1\dots I_l}$, we can contract with the unit vector $n^I \in S^p$ to produce the harmonic $Y^p_l(n)=C_{I_1\dots I_l}n^{I_1}\dots n^{I_l}$. Consider the null vector
\begin{equation}\label{polarization parameterization}
    y_q = (i\hat q, 1),
\end{equation} 
so that $y^2=0$. With this, we can produce the traceless symmetric tensor $y_{qI_1}\dots y_{qI_l}$, which gives rise to the spherical harmonic $Y_l(n) = \sqrt{A_l}(y_q\cdot n)^l$. $A_l$ is a normalization factor 
\begin{equation}
    A_l= \frac{\Gamma(l+\frac{p+1}2)}{2^{1-l}\pi^{\frac{p+1}2}\Gamma(l+1)R^{p}},
\end{equation}
chosen so that
\begin{equation}
   \sqrt{A_{l_1}A_{l_2}}\int_{S^p} d^pn (y_1 \cdot n )^{l_1}(y_2 \cdot n)^{l_2} = \delta_{l_1l_2}(y_1 \cdot y_2)^{l_1}.
\end{equation}
The integral can be evaluated using the moment integration formula (\ref{sphere moment integral}).

In the flat space limit, this harmonic reduces to a plane wave moving in the direction $\hat q$. To see this, we begin by choosing coordinates such that $n = (\sin(\theta)\hat r, \cos (\theta))$. Hence, we have $(y_q \cdot n) ^l=(\cos(\theta) + i\hat q \cdot \hat r\sin (\theta))^l$. Now note that $|(y_q\cdot n)|^2 = \cos^2 (\theta)+(\hat r \cdot \hat q )^2 \sin^2(\theta) \le 1$. Thus, $(y \cdot n)^l \to 0$ as $l \to \infty$ unless $|y_q\cdot n|=1$, which happens only when $\theta = 0, \pi$, or $|\hat r \cdot \hat q| = 1$. Thus, the function goes to zero everywhere except the north and south poles and the great circle determined by $\hat q$. Expanding in $\theta = r/R$ at the north pole, we have 
\begin{equation}
    \lim _{l\to \infty}(y_q \cdot n) ^l= \lim _{l\to \infty}\left(1 + i\hat q \cdot \hat r\frac r R \right)^l = \lim _{l\to \infty}\left(1 + \frac{i\vec q \cdot \vec r}{l} \right)^l =e^{i\vec q \cdot \vec r}.
\end{equation}
Since this harmonic also reduces to a plane wave in the flat limit, we can eliminate the sums over spherical harmonics $Y_{k\alpha}^{*p-1}$ in favour of the auxiliary polarization $y_q$, 
\begin{equation}
    \lim_{l\to \infty }\sum_{k\alpha} Y^{*p-1}_{k\alpha} (\hat q)\OO^\I \propto \lim_{l\to \infty }\OO^l(x, y) = \lim_{l\to \infty }\OO^{I_1 \dots I_l}y_{qI_1}\dots y_{qI_l},
\end{equation}
leading to a considerable simplification for the scattering state (the simplest way to fix the overall normalization is via the two point function, which will be done below). This seems analogous to how the sum over the harmonics $\Y_{Jm}$ produces a delta function $\delta(\hat x- \hat p)$, which also greatly simplifies the noncompact part of the scattering state. Note that we can now clearly see the parity issue: there is a mirror image of the plane wave at the south pole. Moreover, there is a nonvanishing wave moving along the great circle, even in the flat limit. 

To obtain a plane wave localized only at the north pole, we will smear over a range of angular momenta $\sigma=\Delta l$ to produce a bulk wavefunction of the form 
\begin{equation}
    \sum _{{a} } f(l-{a} )  (y_q \cdot n) ^{a}.
\end{equation}
For example, one choice is a Gaussian smearing $f(l-{a}) =(2\sqrt{\pi}\sigma)^{-1} \exp(-(l-{a})^2/4\sigma^2)$, with the limit $\sigma \to \infty, \sigma/R \to 0$ (which can be achieved by choosing $\sigma = R^\gamma$ with $\gamma \in (0, 1)$). To see that this removes everything except the north pole region, let us focus on the great circle $\hat r =\hat q$ where the spherical harmonic does not vanish and takes the form $e^{il\theta}$. We can formally extend the sum to negative $a$ (which introduces only exponentially small error due to the Gaussian) and use Poisson resummation,
\begin{equation}
    \Psi(\theta)=\sum_{a\in \Z}\frac{1}{2\sqrt{\pi}\sigma}  e^{-\frac{(l-{a})^2}{4\sigma^2}} e^{i{a} \theta } = e^{il\theta}\sum_{k\in \Z}e^{-\sigma^2 (\theta-2\pi k)^2}.
\end{equation}
At the north pole $\theta=0$, this gives 
\begin{equation}
    \Psi(0)=1+2e^{-4\pi^2\sigma^2}+\dots\to1
\end{equation}
as $\sigma \to \infty$. At the south pole, we instead have 
\begin{equation}
    \Psi (\pi)=(-1)^l(2e^{-\pi^2\sigma^2}+\dots) \to0
\end{equation}
as $\sigma\to \infty$. The phases $(-1)^a$ are responsible for suppressing the south pole region. Indeed, for any angle $\theta$ at an angular distance greater than $1/\sigma$ from the north pole, we have 
\begin{equation}
    \Psi(\theta) \sim e^{-\sigma^2\theta^2} \to0.
\end{equation}
Thus, the smeared wavefunction is unaffected in the region near the north pole, but it vanishes everywhere else on the sphere. The requirement $\sigma/R\to 0$ is needed to have a well-defined compact momentum across the smearing range, since then $q\sim (l\pm \sigma)/R\to l/R$.

The normalization is simplest to obtain via the two point function. The momentum states must be normalized so that 
\begin{align}\label{two point check}
    \langle \vec P_1| \vec P_2\rangle &= (2\pi)^{D-1}2\omega\delta^{D-1}(\vec P_1-\vec P_2)=(2\pi)^{d+p}2\omega \delta^d(\vec p _1-\vec p_2)\delta^p(\vec q_1-\vec q_2).
\end{align}
Setting
\begin{equation}
    \langle \OO^{l_1}(x_1, y_1)\OO^{l_2}(x_2, y_2)\rangle = \frac{\delta_{l_1l_2}C_{\Delta_{l_1}}(y_1^* \cdot y_2)^{l_1}}{x_{12}^{2\Delta_{l_1}}} =\frac{\delta_{l_1l_2}C_{\Delta_{l_1}}(1+ \hat q_1 \cdot \hat q_2)^{l_1}}{x_{12}^{2\Delta_{l_1}}},
\end{equation}
where
\begin{equation}
    C_{\Delta_{l}} = \frac{\Gamma(\Delta_l)}{2\pi^{d/2}\Gamma(\Delta_l -d/2+1)},
\end{equation}
we show in appendix \ref{appendix B} that
\begin{equation}\label{eq: momentum state}
    \ket{\vec P }  = \underset{\substack{\LA\to \infty \\ l\to \infty \\ l/\chi\LA=q }}{\lim} \sum _{{a}}f(l-{a}) \int dt e^{i\omega_{{a}} t} N_{{a}\omega}\OO^{a}(t, \hat p , y_{q})\ket{0},
\end{equation}
with 
\begin{equation}
    N_{l\omega}=\LA ^{\frac{d-1}2-\Delta_{l}}\xi _{\omega, \Delta_{l}} |\vec p|^{1-\Delta_{l}}2^{\Delta_{l}}\pi ^{\frac{d-2}2}\Gamma\left(1+\Delta_{l}-\frac d 2 \right)  \left ( \frac{\chi \LA \pi^{\frac{p+1}2}\Gamma(l+p-1)}{2^{l-1}q^{p-1}\Gamma(l+\frac{p-1}2)}\right)^{\frac 1 2}.
\end{equation}
The first part is again the same as the pure AdS case \cite{fitzpatrick2011scatteringstatesadscft, Li_2021}, and the second comes from the sphere. With this formula, we can obtain a generic $D$ dimensional $n$ particle S-matrix from $d$ dimensional CFT correlators: 
\begin{equation}\label{generic S matrix}
    \langle P_{k+1}, \dots, P_n| P_{1}, \dots P_k\rangle = \underset{\substack{\LA\to \infty \\ l_i\to \infty \\ l_i/\chi\LA=q_i }}{\lim} \int \prod _{i}dt_i \sum _{{a}_i}f(l_i-{a}_i)e^{i\eta_i\omega_it_i}N_{{a}_i\omega_i}\langle \OO^{{a}_1}(t_1, \hat p_1,  y_{q_1})\dots \OO^{{a}_n}(t_n, \hat p_n,  y_{q_n}^*)\rangle.
\end{equation}
Here, $\eta_i = \pm 1$ for incoming/outgoing particles. Note that we must conjugate $y$ if the particle is outgoing, which is equivalent to replacing $\vec q_i \to -\vec q_i$. We will use the all incoming convention henceforth in writing delta functions.

\section{Four point functions}\label{Sec: Checks}

We will now perform some basic consistency checks of (\ref{generic S matrix}). We will begin with some scalar interaction in $D$ dimensions, dimensionally reduce it to $d+1$ dimensions, apply our formula (\ref{generic S matrix}), and see whether we obtain the expected $D$ dimensional S-matrix. We focus on four point tree level checks, but the analysis can be simply extended to higher-point tree level amplitudes. Extension to loops seems straightforward in principle, but we will not go through the details of regularization which would come into play.

\subsection{Contact interaction}\label{sec contact check}

Suppose we have a $D$ dimensional bulk interaction $\frac{\lambda}{4!}\Phi^4$. This gives a contribution 
\begin{equation}\label{contact diagram}
    i \lambda (2\pi)^D\delta^{D}\left(\sum_i P_i^M\right) = i \lambda (2\pi)^D \delta^{d+1}\left(\sum_i p_i^\mu\right) \delta^p\left(\sum_i q_i^m\right)
\end{equation}
to the S-matrix. Dimensionally reducing on $S^p$ gives the interaction\footnote{We compress the indices on the modes for readability.}
\begin{equation}
    S_d = \frac{\lambda}{4!}\sum _{l_i}c_{l_1l_2l_3l_4} \int_{AdS} d^{d+1} x \sqrt{-g_{AdS}}  \phi ^{l_1}(x)\phi ^{l_2}(x)\phi ^{l_3}(x)\phi ^{l_4}(x),
\end{equation}
with 
\begin{equation}
    c_{l_1l_2l_3l_4} = \int_{S^p} d^pn Y_{l_1}(n)Y_{l_2}(n)Y_{l_3}(n)Y_{l_4}(n).
\end{equation}

Let us now evaluate the $d$ dimensional CFT four point function with such a bulk interaction. The relevant four point contact Witten diagram has an AdS and sphere contribution. The AdS part gives the usual D-function \cite{D_Hoker_1999}, while the sphere contribution is captured by $c_{l_1l_2l_3l_4}$ with the harmonics $Y_{l}(n)= \sqrt {A_l} (y \cdot n)^l$. Thus, we have 
\begin{align}
    \langle \OO^{l_1}(t_1, \hat p_1,  y_{q_1})\OO^{l_2}(t_2, &\hat p_2,  y_{q_2})\OO^{l_3}(t_3, \hat p_3,  y_{q_3}^*)\OO^{l_4}(t_4, \hat p_4,  y_{q_4}^*)\rangle \nonumber \\&= i\lambda D_{\Delta_{l_1}\Delta_{l_2}\Delta_{l_3}\Delta_{l_4}} (t_1, \hat p_1, t_2 , \hat p_2, t_3, \hat p_3, t_4, \hat p_4)\nonumber\\ &\;\;\;\;\times \prod_{i=1}^4\sqrt{A_{l_i}}\int _{S^p} d^pn (y _{q_1}\cdot n)^{l_1} (y _{q_2}\cdot n)^{l_2} (y _{q_3}^*\cdot n)^{l_3} (y _{q_4}^*\cdot n)^{l_4} .  
\end{align}

In the flat space limit, the D-function, with the AdS part of the scattering kernel (\ref{generic S matrix}) from above, reduces to $(2\pi)^{d+1}\delta^{d+1}(\sum_i p_i^\mu)$ \cite{Komatsu_2020, Li_2021}. We must therefore check whether $c_{l_1l_2l_3l_4}$ gives us the remaining momentum conservation factor $(2\pi)^p\delta^p\left(\sum_i q_i\right)$ in the compact directions. The integral can be evaluated using sphere moment integrals, which will be discussed more in section \ref{Sec: Double Mellin representation}. Here we will take a different route. First, note that the normalization $\sqrt{A_l}$ cancels with the sphere part of the normalization $N_{l\omega}$ in the kernel (\ref{eq: momentum state}) in the large $L$ limit. Next, in the flat space limit we can replace the various harmonics $(y_q\cdot n)^l$ with their limit expressions $e^{i\vec q\cdot \vec r}$. To see this, let us make the convenient choice $\sigma_i= \sqrt{|q|_iR/2}$, i.e., we will use a different smearing for the different particles; nothing significant hinges on this choice, it just yields a simpler expression for the envelope. With this choice, the Gaussian envelope around the north pole takes the rotationally symmetric form $e^{i\vec q \cdot \vec r}e^{-\frac{|q|}{2R}r^2}$, which can be seen by expanding $(y\cdot n)^l$ to second order and including the smearing contribution $e^{-\sigma^2(\hat q \cdot \hat r)^2/R^2}$,
\begin{equation}
  \sum_a f(l-a) (y\cdot n)^a \to e^{i\vec q \cdot \vec r} \exp\left(-\frac{|q|}{2R}(r^2 - (\hat q \cdot \vec r)^2) - \frac{\sigma^2}{R^2}(\hat q \cdot \vec r)^2\right) = e^{i\vec q \cdot \vec r} \exp\left(-\frac{|q|}{2R}r^2 \right).
\end{equation}
Let us split the integration over the sphere into two regions: $\theta \in (0, R^{-\frac 1 2 + \epsilon})$ and $\theta \in (R^{-\frac 1 2 + \epsilon}, \pi)$, where $0<\epsilon<1/6$ (the upper bound is to suppress higher order corrections to the Gaussian in the north pole region). The second region is parametrically outside the $1/\sigma \sim 1/\sqrt{R}$ localization region of the smeared wavefunctions discussed earlier, and thus in the flat space limit its contribution to the sphere integral vanishes. The first region localizes to the north pole, and on this scale the curvature of the sphere can be neglected; hence the north pole contribution becomes 
\begin{equation}
    \lim_{R\to \infty} \int_{|r|=0}^{|r|=R^{\frac 1 2 +\epsilon}} d^p r e^{i\sum_i \vec q \cdot \vec r }e^{-\sum_i |q_i|{r^2}/{2R}}.
\end{equation}
Although the angular coordinate $\theta$ is shrinking, the radial coordinate $r=R\theta\sim R^{1/2}$ diverges. Integrating against a test function allows the use of the dominated convergence theorem,\footnote{The upper bound on the integral can be accommodated by including an indicator function, but nothing in the argument changes.} and in a distributional sense the above integral approaches 
\begin{equation}
    \int_{\R^p} d^p r e^{i\sum_i \vec q _i \cdot \vec r} = (2\pi)^p \delta^p \left( \sum_i \vec q_i\right).
\end{equation}
Thus, we have 
\begin{equation}\label{contact sphere contribution}
   \sum _{{a}_i}f(l_i-{a}_i) \int _{S^p} d^pn\prod_{i} (y_{q_i}\cdot n)^{{a}_i}\to  \int_{\R^p}d^p\vec r e^{i\vec r \cdot (\vec q_1 + \vec q_2 - \vec q_3 - \vec q_4)} =  (2\pi)^p\delta^{p}\left(\sum_i \vec q_i \right).
\end{equation}
We therefore recover the expected amplitude (\ref{contact diagram}). Without the smearing, but with generic non-collinear momenta, the integral would instead localize to the north and south poles, giving two delta functions with a phase difference, $(1+(-1)^{l_\Sigma})(2\pi)^{p}\delta^p(\sum_i \vec q_i)$, where $l_\Sigma= \sum_i l_i$. For even $l_\Sigma$ this would give an extra factor of two, while it would give zero for odd $l_\Sigma$. The smearing eliminates the extra delta function and also deals with collinear momenta, where the integral would instead localize to the common great circle.

\subsection{Derivative interaction}

Consider now a $D$ dimensional interaction of the form $\frac{\lambda}{8}(\partial_M \Phi \partial ^M \Phi)^2$. This gives a contribution
\begin{equation}\label{derivative s matrix}
    i\lambda (2\pi)^D\delta^D\left( \sum P_i\right) \times ((P_1 \cdot P_2)(P_3\cdot P_4) + \text{perms}.) 
\end{equation}
to the S-matrix. The dimensionally reduced action gives rise to three types of terms: derivatives can act purely in the noncompact directions $(\nabla_\mu \phi \nabla^\mu \phi)^2$, purely in the compact directions $(\nabla_m \phi \nabla^m \phi)^2$, or in both directions $(\nabla_\mu \phi \nabla ^\mu \phi)( \nabla_m \phi \nabla^m \phi)$. When they act in the noncompact directions, the D-functions are modified, and the expected noncompact kinematic factors $(p_1\cdot p_2)(p_3\cdot p_4)$ arise; this is perhaps best seen through the Mellin space formalism of 
\cite{Penedones_2011, Fitzpatrick_2011, Fitzpatrick_2012}. Moreover, the sphere contribution is identical to (\ref{contact sphere contribution}) above, recovering momentum conservation in the compact directions. When the derivatives act purely in the compact directions, we retain the contact D-function, but the sphere contribution becomes
\begin{equation}\label{derivative sphere ccontribution}
    \prod_{i=1}^4\sum _{{a}_i}f(l_i-{a}_i)N_{{a}_i\omega_i}\sqrt{A_i}\int_{S^p} d^pn\left(\nabla_m(y _{q_1}\cdot n)^{{a}_1} \nabla^m(y _{q_2}\cdot n)^{{a}_2}\right)\left( \nabla_n(y _{q_3}^*\cdot n)^{{a}_3}\nabla^n (y _{q_4}^*\cdot n)^{{a}_4}\right),
\end{equation}
summing over the combinations of (1234). In the flat space limit, we can again simply take $(y_{q_i} \cdot n)^{l_i} \to e^{i\vec q_i \cdot \vec r_i}$ and $\int _{S^p} \to \int_{\R^p}$. Similarly, the derivatives become $\nabla_{S^p} \to \nabla_{\R^p}$. Hence, we find that (\ref{derivative sphere ccontribution}) reduces to
\begin{equation}
    (2\pi)^p \delta^p\left( \sum_i \vec q_i\right) ((\vec q_1 \cdot \vec q_2) (\vec q_3 \cdot \vec q_4) + \text{perms.}).
\end{equation}
The mixed contributions work identically, with the modified D-functions providing factors of $p_i$ and the derivatives of spherical harmonics providing $\vec q_i$. Summing all of these contributions gives back (\ref{derivative s matrix}).

\subsection{Exchange diagram}\label{sec exchange check}

Finally, consider a $D$ dimensional action of the form $\frac{\lambda}{3!}\Phi^3$. We focus for simplicity on the s-channel contribution, 
\begin{equation}
    -\lambda^2 (2\pi )^D \delta^D \left ( \sum_i P_i\right) \frac{i}{-(P_1+P_2)^2},
\end{equation}
since all other contributions work identically. Again, the $\Phi^3$ term will give rise to various cubic interactions between KK modes in the dimensionally reduced theory, and we must sum over all the allowed intermediate KK modes. Thus, letting $K_\Delta(x|X)$ denote the bulk-boundary propagator and $G_\Delta(X, Y)$ denote the bulk-bulk propagator for a particle dual to an operator of weight $\Delta$, the s-channel contribution to the four point function becomes\footnote{We suppress the normalization factors and the smearing; their role is identical to above.} 
\begin{equation}
    -\lambda^2\int_{AdS} dXdY K_{\Delta_{l_1}}(x_1| X)K_{\Delta_{l_2}}(x_2| X) \left(\sum _{\I}c_{\I}(y_i, l_i)G_{\Delta_l}(X, Y)\right)K_{\Delta_{l_3}}(x_3| Y)K_{\Delta_{l_4}}(x_4| Y),
\end{equation}
with 
\begin{equation}
    c_{\I}(y_i, l_i) = \int _{S^p}d^pn (y _{q_1}\cdot n)^{l_1} (y _{q_2}\cdot n)^{l_2}Y_\I^{*p}(n)\int _{S^p}d^pn'Y_\I^{p}( n') (y _{q_3}^*\cdot n')^{l_3} (y _{q_4}^*\cdot n')^{l_4}.
\end{equation}
Let us focus on one of these vertex factors $\int _{S^p}d^pn (y _{q_1}\cdot n)^{l_1} (y _{q_2}\cdot n)^{l_2}Y_\I^{*p}(n)$. We would expect that in the flat space limit, this reduces to a momentum conserving delta function so that the propagating particle carries momentum $\vec q_1 + \vec q_2$. To see that this is indeed the case, let us use (\ref{flat limit of spherical harmonics}). If we also invert the plane wave expansion (\ref{plane wave explansion}) using the orthogonality of spherical harmonics, we find that 
\begin{equation}
    Y^{p}_\I (\Omega_p) \to \sqrt{\frac{q^{p-1}}R} \frac{{i}^{-k}}{(2\pi)^{p/2}}\int _{S^{p-1}}d^{p-1}\hat q e^{i\vec q \cdot \vec r}Y^{(p-1)}_{k\alpha}(\hat q).
\end{equation}
Hence, we have that this vertex becomes 
\begin{align}
    \frac{(-i)^{-k}}{(2\pi)^{p/2}}\sqrt{\frac{q^{p-1}}{R}}\int d^{p-1}\hat q&\int _{\R^p}d^p \vec r e^{i({\vec q_1 + \vec q_2 - \vec q})\cdot \vec r} Y^{*p-1}_{k\alpha}(\hat q) \nonumber \\&= (-i)^{-k}(2\pi)^{p/2} \sqrt{\frac{q^{p-1}}{R}}\int d^{p-1}\hat q \delta^p(\vec q_1 + \vec q_2 - \vec q) Y^{*p-1}_{k\alpha}(\hat q) \nonumber \\&= (-i)^{-k}(2\pi)^{p/2}\sqrt{\frac{1}{Rq^{p-1}}}\int d^{p-1}\hat q \delta(|q_1+q_2|-q)\delta^{p-1}(\widehat{q_1+q_2}-\hat q)Y^{*p-1}_{k\alpha}(\hat q)  \nonumber \\ &=(-i)^{-k}(2\pi)^{p/2} \sqrt{\frac{1}{Rq^{p-1}}}Y^{*p-1}_{k\alpha}(\widehat{q_1+q_2})\delta(|q_1+q_2|-q).
\end{align}
The other vertex yields a similar conjugated contribution, which cancels the phases. Noting that $\sum_{\I} \to R\int dq \sum_{k\alpha}$, we get 
\begin{align}
    (2\pi)^p\int dq \delta(|q_1+q_2|-q)&\delta(|q_3+q_4|-q)\frac{G_{\Delta_q}(X, Y)} {q^{p-1}} \sum_{k\alpha} Y^{*p-1}_{k\alpha}(\widehat{q_1+q_2)}Y^{p-1}_{k\alpha}(\widehat{q_3+q_4)} \nonumber \\&=(2\pi)^p\frac{\delta(|q_1+q_2|-|q_3+q_4|)\delta^{p-1}(\widehat{q_1+q_2}-\widehat{q_3+q_4})}{|q_1+q_2|^{p-1}}G_{\Delta_{
    |q_1+q_2|}}(X,Y) \nonumber \\&= (2\pi)^{p}\delta^p\left(\sum_i\vec q_i\right)G_{\Delta_{|q_1+q_2|}}(X, Y).
\end{align}
In addition to obtaining overall momentum conservation, we have that the exchanged particle must be in a representation $l$ such that $l/R = q = |q_1+q_2|$. We now need the Mellin result for the flat limit of the exchange diagram for an exchanged particle of mass $m_l$ \cite{Penedones_2011}, which simply gives the expected
\begin{equation}
    (2\pi)^{d+1}\delta^{d+1}\left( \sum_i p^\mu_i\right) \frac{-i\lambda^2}{-(p_1^\mu+p_2^\mu)^2-m_l^2} .
\end{equation}
Using $m_l^2 =l(l+p-1)/R^2 \to q^2$ for the KK mode, we find  
\begin{equation}
    (2\pi)^D\delta^D\left( \sum _{i}P_i\right) \frac{-i\lambda^2}{-(p_1^\mu+p_2^\mu)^2-(\vec q_1+ \vec q_2)^2}.
\end{equation}
We have reconstructed the full $D$ dimensional exchanged momentum.

These calculations suggest how other, more generic diagrams are obtained. The Mellin results for the various D-functions give us the noncompact parts of the amplitude. The harmonics $(y \cdot n)^l$ become plane waves, and the selection rules for integrals of spherical harmonics enforce strict momentum conservation at the vertices in the flat space limit. Internal propagators carrying KK modes of mass $m_l$ are reinterpreted as particles carrying compact momentum $l/R$. We can see that sums over the exchanged KK levels $\sum_l$ become integrals over the compact momenta, $R\int dq$, suggesting how loop integrals will arise. 

Of course, the above checks are not terribly surprising. They are especially not surprising since we computed the CFT correlation function using Witten diagrams, meaning the flat space limit was fairly apparent from the start. However, they have hopefully given evidence that (\ref{generic S matrix}) is precisely the right object to compute a generic flat space S-matrix from CFT data.

\section{Double Mellin representation}\label{Sec: Double Mellin representation}

As mentioned above, a particularly nice way to go from AdS correlation functions to S-matrices is via the Mellin space formalism of \cite{Penedones_2011,Fitzpatrick_2011}, which was proven for massless and massive cases using the scattering kernel in \cite{Fitzpatrick_2012, Li_2021}. We first write the (connected part of the) CFT correlation function in Mellin space,
\begin{equation}
    \left \langle \prod_{i=1}^n \OO^{l_i}(x_i) \right \rangle_c = \int [d\delta_{ij}] \prod_{i<j} \Gamma(\delta_{ij}) (x_{ij})^{-\delta_{ij}} \M(\delta_{ij}),
\end{equation}
where $\sum_{j\neq i}\delta_{ij}=\Delta_{i}$, and in our coordinates $x_{ij}=2(\cos(\tau_1-\tau_2) -\hat x_1 \cdot \hat x_2)$. Penedones' idea was that a suitable integral transform of the Mellin space correlator $\M(\delta_{ij})$ would give the S-matrix in the flat space limit, with the Mellin variables $\delta_{ij}$ being related to the Mandelstam invariants $s_{ij}=-(p_i^\mu + p_j^\mu)^2$. More precisely, Penedones conjectured that the connected part of the S-matrix $T(s_{ij})$ for massless particles is given by
\begin{equation}\label{Penedones formula}
    T(s_{ij}) =\lim_{\LA \to \infty} (2\pi)^{d+1}\delta^{d+1}\left( \sum_i p_i^\mu \right) \frac 1 \N \int_{-i\infty}^{i\infty}  d\alpha e^\alpha \alpha^{\frac 1 2(d -\sum_i \Delta_i)}\M\left(\delta_{ij} = -\frac{\LA^2}{4\alpha}s_{ij}\right),
\end{equation}
where $\N$ is a normalization factor 
\begin{equation}
    \N=\frac{\pi^{\frac d 2 }\LA^{\frac{n(1-d)}{2}+d+1}}{2}\prod_{i=1}^n \frac{C_{\Delta_i}}{\Gamma(\Delta_i)}.
\end{equation}
The formula was proven using the pure AdS part of the scattering kernel (\ref{generic S matrix}). Doing a saddle point integration on the Mellin variables $\delta_{ij}$ and doing the integration over $t_i$ precisely gives Penedones' formula \cite{Fitzpatrick_2012,Li_2021}. 

It is natural to wonder if there is a similar formula to extract the S-matrix with generic external kinematics in the flat limit via a `double Mellin representation' which incorporates the sphere variables. Indeed, this has been anticipated in some form in \cite{Rastelli:2016nze,Rastelli_2018,Aprile_2020,Aprile_2021}, and most explicitly in \cite{Aprile_2023}. Including the sphere part of the smearing kernel in (\ref{generic S matrix}) allows us to verify that this double Mellin representation does give the flat space S-matrix, including the contributions to the Mandelstam invariants from the compact momenta. We derive the general result in section \ref{general double mellin result} and check it for the four point exchange diagram in section \ref{recheck of exchange diagram}. 

\subsection{General formalism}\label{general double mellin result}

Let us write the CFT correlation function in double Mellin space \cite{Aprile_2023}.\footnote{In \cite{Aprile_2023}, the authors pull out a kinematic factor in front, but we will not do this.} Denoting the noncompact Mellin variables as $\hat \delta_{ij}$ and the compact ones as $\check \delta_{ij}$, we have
\begin{equation}\label{Double mellin rep}
     \left \langle \prod_{i=1}^n \OO^{l_i}(x_i, y_i) \right \rangle_c = \int [d \hat \delta_{ij}] \sum_{\ch}\prod_{i<j} \frac{\Gamma(\hat \delta_{ij}) (y_{ij})^{\check \delta_{ij}}} {\Gamma(\check \delta_{ij}+1)(x_{ij})^{\hat \delta_{ij}}}\M(\hat \delta_{ij}, \check \delta_{ij}),
\end{equation}
with $y_{ij}=y_i \cdot y_j$ and the constraints 
\begin{equation}
    \sum_{j\neq i}\hat \delta_{ij} = \Delta_{l_i}, \;\;\sum _{j\neq i} \check \delta_{ij} = l_i.
\end{equation}
We will stop explicitly writing the subscript $y_q$ on the null polarizations, but they should be understood to be parameterized as in (\ref{polarization parameterization}). The form of the compact Mellin part is suggested by the integrals from above, 
\begin{equation}\label{contact sphere integral}
    \int _{S^p} d^pn \prod_i (y_i \cdot n)^{l_i}.
\end{equation}
These integrals can be evaluated by making use of the sphere moment integration formula \cite{lee1998threepointfunctionschiraloperators},
\begin{equation}\label{sphere moment integral}
    \int_{S^p} d^pn n^{I_1}\dots n^{I_{2m}}=\frac{2\pi^{(p+1)/2}R^p}{2^m\Gamma(m+\frac{p+1}2)}\sum_{\text {pairs}} \delta_{I_{a_1}I_{a_2}}\dots\delta_{I_{a_{2m-1}}I_{a_{2m}}}.
\end{equation}
Accounting for the ways of combining $y_i$ with $y_j$ yields 
\begin{equation}\label{contact Mellin representation}
    \frac{2\pi^{(p+1)/2}R^p \prod_i l_i!}{2^{l_\Sigma/2}\Gamma(\frac{l_\Sigma+p+1}2)} \sum_{\ch}\prod _{i<j} \frac{y_{ij}^\ch}{\Gamma(\ch+1)},
\end{equation}
with precisely the constraint $\sum_{j \neq i}\ch=l_i$. Here, $l_\Sigma = \sum_i l_i$. For convenience we define
\begin{equation}
    C(l_i)=\frac{2\pi^{(p+1)/2}R^p \prod_i l_i!}{2^{l_\Sigma/2}\Gamma(\frac{l_\Sigma+p+1}2)}.
\end{equation}
This prefactor will be absorbed into the normalization. Recall also that the sphere part of the full S-matrix (\ref{generic S matrix}) contains the normalization  
\begin{equation}
    \left ( \frac{\chi \LA \pi^{\frac{p+1}2}\Gamma(l+p-1)}{2^{l-1}q^{p-1}\Gamma(l+\frac{p-1}2)}\right)^{\frac 1 2},
\end{equation}
which is required to cancel the factors of $\sqrt{A_l}$ coming from our choice of normalization for the spherical harmonics. This will also be absorbed into the total normalization below.  

Plugging (\ref{Double mellin rep}) into (\ref{generic S matrix}) gives an expression for the flat space S-matrix in terms of $\M(\ha, \ch)$ with integrations over $\ha$, $t_i$, sums over $\ch$, and the smearing sum. The integrals and sums can be evaluated by saddle point. We must use the massive version of Penedones' formula \cite{Paulos_2017, Komatsu_2020}, proven using the scattering kernel in \cite{Li_2021}, since the nonzero compact momenta $q$ arise from the masses $m_l$. This shifts $\hat s_{ij}\to \hat s_{ij}-(m_i+m_j)^2$ in (\ref{Penedones formula}), where $m_i$ are the masses of the particles after taking the flat limit, $\Delta_i=Lm_i$. Moreover, it allows a saddle point evaluation of the integral over $\alpha$ \cite{Li_2021}, giving an even simpler result 
\begin{align}
    T(s_{ij}) =&\lim_{\LA \to \infty} (2\pi)^{d+1}\delta^{d+1}\left( \sum_i p_i^\mu \right) L^{\frac{n(1-d)}{2}+d+1}  \M\left(\ha = -\frac{\LA^2}{2\Delta_\Sigma}(\sha-(m_i+m_j)^2) \right).
\end{align}
Here, $\Delta_\Sigma = \sum_i\Delta_i$. Note that in the large $L$ and $l_i$ limit, $\Delta_i = l_i/\chi$ and $m_i=l_i/(\chi L)=|q_i|.$

The AdS part of the computation is identical to what has been done before. As such we will focus on the sum over $\ch$. Since we will be interested in scaling the Mellin variables $\ch$ to large values, we can use Stirling's approximation on $\Gamma(\ch+1)$. Moreover, we can introduce Lagrange multipliers to enforce the constraint $\sum_{j\neq i}\ch = l_i$. Hence, the sum becomes 
\begin{equation}\label{Eq: stirling Mellin sum}
    \int \frac{d\lambda_i}{(2\pi)^n}\sum_\ch \prod_{i<j} \frac{1}{\sqrt{2\pi\ch}}\exp\left[ \ch\ln(y_{ij})+\ch-\ch\ln\ch -i\sum_i \lambda_i \left(l_i- \sum_{j\neq i}\ch\right)\right].
\end{equation}
As in \cite{Li_2021}, it is convenient to change variables $\beta_i = e^{i\lambda_i}$, and the exponent becomes 
\begin{equation}
    \exp\left[ \sum_{i<j}\left(\ch \ln(y_{ij})+\ch -\ch\ln\ch \right) -\sum_i l_i\ln\beta_i + \sum_{i<j}\ln(\beta_i\beta_j) \ch\right].
\end{equation}
The saddle point equation for $\ch$ forces $\ch^*=\beta_i\beta_j y_{ij}$. Plugging back in gives the exponent
\begin{equation}
    \exp\left[ \sum_{i<j}\beta_i \beta_j y_{ij}-\sum_i l_i \ln\beta_i\right].
\end{equation}
Evaluating this by saddle point then requires
\begin{equation}
    \beta_i \sum_j \beta_jy_{ij}=l_i.
\end{equation}
Assuming compact momentum is conserved $\sum_i \vec q_i = 0$, we can solve this with $\beta_i^* = l_i/\sqrt{l_\Sigma}$. The saddle value of $\ch$ is then
\begin{equation}
    \ch^* = \frac{l_il_j}{l_\Sigma}y_{ij}=\frac{(l_i+l_j)^2+\chi^2L^2\check s_{ij}}{2l_\Sigma} =\frac{\chi^2L^2}{2l_\Sigma}(\check s_{ij}+(|q_i|+|q_j|)^2),
\end{equation}
where $\check s_{ij}=-(\vec q_i+\vec q_j)^2$. Thus, including the sphere contribution in Penedones' formula gives us 
\begin{align}\label{sphere Penedones formula}
    T(s_{ij}) =&\lim_{\LA \to \infty} \frac{(2\pi)^D}{ \N'}\delta^D\left( \sum_i P_i\right) \M\left(\ha = -\frac{\LA^2}{2\Delta_\Sigma}(\sha-(m_i+m_j)^2), \ch = \frac{\chi^2 L^2}{2l_\Sigma}(\sh+(|q_i|+|q_j|)^2)\right).
\end{align}
With our conventions, we need to modify the normalization to 
\begin{equation}\label{Mellin normalization}
    \N' = L^{\frac{n(d-1)}{2}-d-1} \times C(l_i) \prod_{i=1}^n \left ( \frac{\chi \LA \pi^{\frac{p+1}2}\Gamma(l_i+p-1)}{2^{l_i-1}q^{p-1}\Gamma(l_i+\frac{p-1}2)}\right)^{-\frac 1 2},
\end{equation}
which accounts for the prefactor in (\ref{contact Mellin representation}) and our normalization of the two point function. It is possible to undertake a more precise analysis accounting for the Gaussian determinants around the saddle points and allowing for a momentum nonconservation, but it seems that this is not necessary for the same reasons as discussed in appendix C.1 of \cite{Li_2021}. In particular, the compact Mellin structure (\ref{contact Mellin representation}) arises from the sphere moment integrals (\ref{sphere moment integral}) with an overall normalization factor. In the S-matrix, we need a smeared version of the moment integral, and as we saw in section \ref{sec contact check}, this reduces to the momentum conserving delta function $(2\pi)^p \delta^p \left( \sum_i \vec q_i \right)$ in the flat space limit. Being slightly more precise, we assume that the fluctuations around the saddle are small and that the Mellin amplitude in the saddle region is sufficiently smooth across this range so that it can be pulled out of the compact Mellin sum (and smearing sum),
\begin{equation}
    \lim_{L\to \infty} \sum_{a_i}f(l_i-a_i)C(a_i)\sum _\ch \prod _{i<j}\widetilde\M(\ha, \ch)  \frac{y_{ij}^\ch}{\Gamma(\ch+1)}=\lim_{L\to \infty}\widetilde \M(\ha, \ch^*)\sum_{a_i}f(l_i-a_i)C(a_i)\sum _\ch \prod _{i<j} \frac{y_{ij}^\ch}{\Gamma(\ch+1)}.
\end{equation}
Here, $\widetilde \M$ is the Mellin amplitude with the overall sphere normalizations removed, $\widetilde \M = \M/\N'$. The remaining sum is the same as the contact diagram (\ref{contact sphere integral}) with smearing, and this reduces to $(2\pi)^p \delta^p \left( \sum_i \vec q_i \right)$ as in section \ref{sec contact check}. The more complicated analysis in \cite{Fitzpatrick_2012} is needed for $d$ dimensional massless particles, for which the saddle point equations are not fully determined and the integral over $\alpha$ remains.

\subsection{Exchange diagram revisited}\label{recheck of exchange diagram}

We will now check (\ref{sphere Penedones formula}) for the case of the exchange diagram. We will perform the saddle point on $\ha$ first and then evaluate the saddle point over $\ch$, i.e. we will approximate $\M(\ha, \ch) \approx \M(\ha^*, \ch)$. This has the effect of replacing the complicated Mellin representation of the exchange by the propagator $i\left[\hat s - \frac{l(l+p-1)}{(\chi L)^2}\right]^{-1}$, where $\hat s$ is the noncompact Mandelstam invariant $-(p_1^\mu+p_2^\mu)^2$. This replacement is only justified upon taking the large $L$ limit, where subleading errors are suppressed. A more systematic way to find the exact $\M(\ha, \ch)$ is by using the higher dimensional version of the spherical harmonic addition theorem; this allows us to replace the sums $\sum_{k\alpha}Y^*_{lk\alpha}(n)Y_{lk\alpha}(n')$ from the KK exchange by a Gegenbauer polynomial in $n\cdot n'$. One can then evaluate the sphere integrals by making use of the moment integral formula. However, the resulting expressions are very complicated and unwieldy. Taking the simplified route, we will instead be interested in expanding\footnote{We omit the overall $-\lambda^2$ in the intermediate expressions.}
\begin{equation}
    I=\int_{S^p} d^pnd^pn' (y_1\cdot n)^{l_1} (y_2\cdot n)^{l_2} (y_3^*\cdot n')^{l_3} (y_4^*\cdot n')^{l_4}\sum_{lk\alpha}\frac{iY^*_\I(n)Y_\I(n')}{\hat s-l(l+p-1)/(\chi L)^2}
\end{equation}
in power of $(y_i \cdot y_j)^{\ch}$ and extracting the coefficient $\M(\ch)$. Here we have canceled the normalizations $\sqrt{A_l}$ with the product in the normalization (\ref{Mellin normalization}). First note that $\left[\hat s - \frac{l(l+p-1)}{(\chi L)^2}\right]^{-1}=\left[\hat s +\nabla^2_{S^p}\right]^{-1}$ acting on ${Y^*_{lk\alpha}}(n)$. By integrating by parts, we can transfer the operator to act on $(y_1 \cdot n)^{l_1}(y_2\cdot n)^{l_2}$. This simplifies the sum $\sum_{lk\alpha}Y^*_{lk\alpha}(n)Y_{lk\alpha}(n')=\delta^p(n-n')$, which eliminates the integral over $n'$. To determine the action of the operator $\left[\hat s +\slap\right]^{-1}$ on $(y_1 \cdot n)^{l_1}(y_2\cdot n)^{l_2}$, we first introduce some notation:
\begin{align}
    F_{k}&= (y_1\cdot n)^{l_1-k} (y_2\cdot n)^{l_2-k} \\m_k^2&=\frac{(l_1+l_2-2k)(l_1+l_2-2k+p-1)}{(\chi L)^2}.
\end{align}
The action of $\slap$ on $F_k$ is 
\begin{equation}
    \slap F_k = -m_k^2F_k+\frac{2(l_1-k)(l_2-k)y_{12}}{(\chi L)^2}F_{k+1}.
\end{equation}
Note that $F_0= (y_1\cdot n)^{l_1}(y_2\cdot n)^{l_2}$. Let us use the ansatz
\begin{equation}\label{laplacian inversion}
    \left[\hat s +\slap\right]^{-1}F_0=\sum_{k=0}^{\min(l_1, l_2)}c_k F_k.
\end{equation}
Applying $\left[\hat s +\slap\right]$ to both sides and using the recursion yields 
\begin{align}
    F_0 = c_0(\hat s-m_0^2)F_0 + \sum_{k=1}^{\min(l_1, l_2)}\left[ c_k(\hat s-m_k^2) + c_{k-1}\frac{2(l_1-k+1)(l_2-k+1)y_{12}}{(\chi L)^2}\right]F_k.
\end{align}
Hence, we have that 
\begin{align}
    c_0&=\frac 1{\hat s-m_0^2} \\ c_{k}&=-\frac{2(l_1-k+1)(l_2-k+1)y_{12}}{(\chi L)^2(\hat s-m_k^2)}c_{k-1}.
\end{align}
Solving this recursion gives 
\begin{equation}
    c_k=\left( \frac{-2y_{12}}{(\chi L)^2}\right)^k \frac{l_1!l_2!}{(l_1-k)!(l_2-k)!}\prod_{r=0}^k \frac 1{\hat s-m_r^2}.
\end{equation}
Thus,
\begin{equation}
    I = i\sum_{k=0}^{\min(l_1, l_2)} \left( \frac{-2y_{12}}{(\chi L)^2}\right)^k \frac{l_1!l_2!}{(l_1-k)!(l_2-k)!}\prod_{r=0}^k \frac 1{\hat s-m_r^2} \int_{S^p} d^pn (y_1\cdot n)^{l_1-k}(y_2\cdot n)^{l_2-k}(y_3^*\cdot n)^{l_3}(y_4^*\cdot n)^{l_4}.
\end{equation}
This is simply the previous integral (\ref{contact sphere integral}) with the levels shifted by $-k$. Shifting the sum and normalizations by $k$ and swapping the sum over $k$ with the sum over $\ch$, we have
\begin{equation}\label{exchange mellin representation}
    I = C(l_i) \sum_{\ch} \M(\ch) \prod _{i<j} \frac{y_{ij}^\ch}{\Gamma(\ch+1)},
\end{equation}
with 
\begin{equation}
   \M(\ch)=i\sum_{k=0}^{\check \delta_{12}} \left( \frac{-4}{(\chi L)^2}\right)^k \frac{\Gamma(\check \delta_{12}+1)}{\Gamma(\check \delta_{12}-k+1)} \frac{\Gamma(\frac{l_\Sigma +p+1}2)}{\Gamma(\frac{l_\Sigma +p+1}2-k)}\prod _{r=0}^k \left [ \hat s-\frac{(l_1+l_2-2r)(l_1+l_2-2r+p-1)}{(\chi L)^2}\right]^{-1}.
\end{equation}
For large $l_i, L, \ch$, evaluating at the saddle point yields 
\begin{align}
    \M(\ch^*)&\to \sum_{k=0}^\infty \frac{i}{\hat s-(|q_1|+|q_2|)^2}\left[ \frac{-2\check \delta_{12}^*l_\Sigma/(\chi L)^2}{\hat s-(|q_1|+|q_2|)^2}\right]^k \nonumber \\ &=\frac{i}{\hat s-(|q_1|+|q_2|)^2 + 2\check \delta_{12}^*l_\Sigma/(\chi L)^2}\nonumber \\&=\frac{i}{\hat s+\check s} =\frac i S,
\end{align}
where $\check s = -(\vec q_1+\vec q_2)^2$ and $S$ is the $D$ dimensional s-channel Mandelstam invariant $S=\hat s + \check s$. The prefactor in (\ref{exchange mellin representation}) cancels with the normalization (\ref{Mellin normalization}), and we again recover the $D$ dimensional exchange diagram as in section \ref{sec exchange check} upon restoring the overall $-\lambda^2$. Note that the sum (upon taking the limit) only converges when $|\hat s - (|q_1|+|q_2|)^2| > |\check s +(|q_1|+|q_2|)^2|$, which holds if the compact momenta $|q_i|$ are small enough relative to the center of mass energy $S$. Outside of this regime, we must analytically continue the limiting expression.

\section{Conclusion}\label{Sec: Conclusion}

In this article we have described a way to construct generic scattering states for scalar fields in the flat space limit of AdS/CFT (\ref{eq: momentum state}). This has allowed us to extract the flat space S-matrix from CFT correlation functions in two different ways, by using either the direct formula (\ref{generic S matrix}) or the Mellin space formula (\ref{sphere Penedones formula}). We have performed some basic checks of the formula for simple scalar field interactions at tree level. 

One potential application is in Carrollian holography. The formula for the momentum state (\ref{eq: momentum state}) gives us an expression for the creation operator $a^\dagger(\vec P)$ in terms of a $d$ dimension CFT operator, which is in turn related to the $D-1$ dimensional Carrollian field $\boldsymbol{O}(u, \hat P)$ by \cite{kulkarni2025carrolliancelestialcorrelatorsgeneral}
\begin{equation}
    \boldsymbol{O}(u, \hat P) =\int _0^\infty \frac{d\omega}{2\pi}(i\eta\omega)^{\frac{D-4}{2}}a^\eta (\vec P)e^{i\eta \omega u-\varepsilon\omega}.
\end{equation}
Here, $\eta =\pm 1$ distinguishes incoming and outgoing particles, $a^ +=a, a^- =a^\dagger$, and $\varepsilon$ is a regulator. Note that $\hat P = (\cos(\alpha) \hat p, \sin (\alpha) \hat q)$, where $\tan (\alpha) = |q|/|p|$. We see that the additional angular direction which mixes the AdS and sphere directions in the flat space limit arises as the ratio of compact to noncompact momentum. This formula in principle allows us to turn S-matrices into Carrollian CFT correlation functions; hence, if (\ref{generic S matrix}) gives us the higher dimensional S-matrix, it also allows us to turn $d$ dimensional CFT correlation functions into $D-1$ dimensional Carrollian CFT correlation functions,
\begin{align}
    \langle \boldsymbol{O}(u_1, \hat P_1)\dots \boldsymbol{O}(u_n, \hat P_n)\rangle &\stackrel{?}{=}\int \prod_{i=1}^n\frac{d\omega_i}{2\pi}(i\eta_i\omega_i)^{\frac{D-4}2}e^{i\eta_i\omega_i u_i-\varepsilon\omega_i}\nonumber \\ &\times \underset{\substack{\LA\to \infty \\ l_i\to \infty \\ l_i/\chi\LA=q_i }}{\lim} \int \prod _{i=1}^ndt_i \sum _{{a}_i}f(l_i-{a}_i)e^{i\eta_i\omega_it_i}N_{{a}_i\omega_i}\langle \OO^{{a}_1}(t_1, \hat p_1,  y_{q_1})\dots \OO^{{a}_n}(t_n, \hat p_n,  y_{q_n}^*)\rangle.
\end{align}
Here, $\boldsymbol{O}(u, \hat P)$ represents the Carrollian field dual to the massless bulk scalar $\Phi(X)$, and $\OO$ are the CFT operators dual to the KK modes of $\Phi(X)$.  The simple normalization (\ref{two point check}) indeed gives the $D-1$ dimensional Carrollian two point function, as expected from $D$ dimensional momentum conservation; this also implies that the Carrollian two point Ward identity is satisfied. Thus, it seems that we might obtain a $D-1$ dimensional Carrollian field theory from adjoining the large R-symmetry charge sector with the Carrollian limit of the CFT$_d$ rather than a $d$ dimensional theory. If this is correct, what is happening is not a KK decompactification in the boundary theory. It is more subtle: the compact R-symmetry group $SO(p+1)$ is becoming the noncompact $ISO(p)$, and it seems to gain an interpretation as a spacetime direction via the polarizations $y_I$ and the angle $\alpha$. By this we mean that $ISO(p)$ seems to be part of the spacetime algebra rather than being an internal symmetry.

Other possible applications include extracting unrestricted $D$ dimensional S-matrices from knowledge of higher point CFT correlation functions. One could hope to undertake an analysis like \cite{Chester_2018} for higher point functions or repeat the analysis for four point functions with generic kinematics as a consistency check. Tree level four point functions for $1/2$ BPS modes with arbitrary external KK charge are known \cite{Alday_2021}. However, to make use of this would require understanding higher spin fields rather than just scalars. Indeed some new issues seem to arise when looking at more complicated dimensional reductions like in supergravity. Higher spin fields can have polarizations in the compact and noncompact directions, and in general the KK modes are in more complicated representations of $SO(p+1)$ corresponding to vector and tensor spherical harmonics \cite{lee1998threepointfunctionschiraloperators}. Moreover, modes can `mix', meaning the dimensionally reduced fields can really be combinations of modes coming from distinct higher dimensional fields. Some of the difficulties with taking the flat space limit of massive higher spin fields which become massless are discussed in \cite{berenstein2025spinningfieldslorentzianads}, particularly the subtleties of how massive polarizations are supposed to become massless polarizations. It would be interesting to understand these complications in more detail.

\section*{Acknowledgments}
I am grateful to James Read and Emil Have for discussions. ChatGPT 6 assisted with the computations in section \ref{recheck of exchange diagram} and suggested the smearing to remove the south pole. This work was completed with support from the Rokos internship.

\appendix
\section{Flat limit of spherical harmonics}\label{Appendix A}

The Laplacian on $S^p$ is 
\begin{equation}
    \nabla ^2_{S^p} = \frac 1 {R^2}\left(\dd_\theta^2 + (p-1) \cot\theta \dd_\theta + \csc^2\theta \nabla^2_{S^{p-1}}\right).
\end{equation}
Separating variables, we can write the spherical harmonics $Y^p_\I$ as 
\begin{equation}
    Y^p_{lk\alpha}(\theta, \Omega_{p-1}) = f^l_k (\theta) Y^{p-1}_{k\alpha}(\Omega_{p-1}),
\end{equation}
where 
\begin{align}
    \hat \nabla^2_{S^p}Y^p_{lk\alpha} &= -l(l+p-1)Y^p_{lk\alpha} \\ \hat \nabla^2_{S^{p-1}}Y^{p-1}_{k\alpha} &= -k(k+p-2)Y^{p-1}_{k\alpha} \\{f^l_k}''+(p-1) \cot\theta {f^l_k}'&+\left[l(l+p-1) - \frac{k(k+p-2)}{\sin^2\theta}\right]{f^l_k}=0.
\end{align}
The solutions to this last equation can be written in terms of Gegenbauer polynomials, yielding the result 
\begin{equation}
    f^l_k(\theta)  \propto  \sin^k(\theta)C_{l-k}^{k+(p-1)/2}(\cos(\theta)).
\end{equation}
The flat-space limit of this equation can be obtained by substituting $r = R\theta$, $R\to \infty$, $\theta \ll 1$, $l\to \infty$ with $l/R := q$ fixed. This zooms in to the north pole of $S^p$, and the angular momentum $l$ becomes the flat space momentum $q$. The radial equation becomes that of cylindrical coordinates in flat space,\footnote{Here, prime denotes derivative with respect to $r$.}
\begin{equation}
    f''_k+ \frac{p-1}{r}f'_k+ \left( q^2 -\frac{k(k+p-2)}{r^2}\right)f_k=0.
\end{equation}
Setting $\nu  =p/2-1$, this has solution
\begin{equation}
    f_k(r) \propto\frac{J_{k+\nu}(qr)}{(qr)^\nu }.
\end{equation}
It follows that $f^l_k(\theta) \to f_k(r)$ in the flat space limit. This can also be obtained using the Mehler-Heine limit of the Gegenbauer polynomials \cite{DLMF18.11ii}, and the precise normalization is obtained by normalizing the $p$ and $p-1$ dimensional spherical harmonics. We finally have
\begin{equation}
    Y^p_{lk\alpha}(\theta, \Omega_{p-1}) \underset{\substack{R\to \infty \\ l\to \infty \\ l/R=q }}{\longrightarrow} \sqrt{\frac{q^{p-1}}{R}}\frac{J_{k+\nu} (qr)}{(qr)^{\nu}} Y^{p-1}_{k\alpha}(\Omega_{p-1}).
\end{equation}

\section{Two point normalization}\label{appendix B}

In this appendix we verify the normalization of the scattering kernel (\ref{eq: momentum state}). Firstly, we note that the AdS part of the computation is identical to that in appendix B of \cite{Li_2021}: the noncompact factor from each mode at level ${a}$ gives $(2\pi)^d 2\omega_{a}\delta^d{(\vec p_1 - \vec p_2})$, since orthogonality of spherical harmonics forces the incoming and outgoing states to have the same level ${a}$. The only subtlety is due to the smearing; however, since we are only summing over $l$ that give rise to the same $q$, $\omega_{a}$ is roughly constant, and it can safely be pulled out of the smearing sum, along with the noncompact delta function. The remaining sphere contribution becomes 
\begin{equation}
    \underset{\substack{\LA\to \infty \\ l_i\to \infty \\ l_i/\chi\LA=q_i }}{\lim} \sum _{{a} } f(l_1-{a})f(l_2 - {a}) \left ( \frac{2\pi^{\frac{p+1}2}R\Gamma({a}+p-1)}{q_{a}^{p-1}\Gamma({a}+\frac{p-1}2)}\right) \left(\frac{1+\hat q_1 \cdot \hat q_2}{2} \right)^{a}.
\end{equation}
Since ${a}$ is forced to be large, and since $q_{a}$ is roughly constant across the smearing range, we can replace the second part with its large ${a}$ limit. It becomes 
\begin{equation}
    \left ( \frac{2\pi^{\frac{p+1}2}R\Gamma({a}+p-1)}{q_{a}^{p-1}\Gamma({a}+\frac{p-1}2)}\right) \left(\frac{1+\hat q_1 \cdot \hat q_2}{2} \right)^{a} \to (2\pi)^p \frac{R}{q_{a}^{p-1}} \delta^{p-1}(\hat q_1 - \hat q_2).
\end{equation}
This can be obtained by choosing coordinates on $S^{p-1}$ such that $\hat q_1 \cdot \hat q_2=\cos (\theta)$ and then normalizing the integration over $S^{p-1}$. The factor $1/q_{a}^{p-1}$ is required to convert the spherical delta function $\delta(q_1-q_2)\delta^{p-1}(\hat q_1 - \hat q_2)$ to the Cartesian $\delta^p(\vec q_1 - \vec q_2)$; it can also be pulled out of the sum, since $q_{a}$ is roughly constant. Thus, it remains to show that 
\begin{equation}
    \underset{\substack{R\to \infty  \\\sigma \to \infty \\\sigma/R\to 0}}{\lim}R\sum _{a} f(Rq_1- {a}) f(Rq_2- {a}) = \delta(q_1-q_2).
\end{equation}
For the Gaussian smearing $f(l-{a})=(2\sqrt{\pi}\sigma)^{-1}\exp{(-(l-{a})^2/4\sigma^2)}$, we indeed find
\begin{align*}
    \underset{\substack{R\to \infty   \\\sigma \to \infty \\\sigma/R\to 0}}{\lim}  R\sum _{{a}}\frac{1}{4\pi \sigma^2}e^{-\frac{(Rq_1-{a})^2}{4\sigma^2}}e^{-\frac{(Rq_2-{a})^2}{4\sigma^2}} &\approx \underset{\substack{R\to \infty\\\sigma \to \infty \\\sigma/R\to 0}}{\lim} R\int_{-\infty}^\infty d{a} \frac{1}{4\pi \sigma^2}e^{\frac{-(Rq_1-{a})^2-(Rq_2-{a})^2}{4\sigma^2} } \\&=\lim_{\sigma/R\to 0}  \frac{R}{\sqrt{8\pi}\sigma}e^{-\frac{R^2(q_1-q_2)^2}{8\sigma^2}} \\&=\delta(q_1-q_2).
\end{align*}

\bibliographystyle{JHEP}
\bibliography{refs.bib}

\end{document}